# Temperature dependence of in-plane correlation lengths in exchange biased Co/FeF$_2$

X. Lu[1], S. Roy[2*], E. Blackburn[3], Mikhail Erekhinsky[1], Ivan K. Schuller[1], J. B. Kortright[4] and S. K. Sinha[1]

[1] Physics Dept. University of California San Diego, La Jolla CA 92093
[2] Advanced Light Source, Lawrence Berkeley National Laboratory, Berkeley CA 94720
[3] School of Physics and Astronomy, University of Birmingham, Birmingham, B15 2TT, UK.
[4] Material Sciences Division, Lawrence Berkeley National Laboratory, Berkeley CA 94720

## Abstract

We have measured resonant soft x-ray diffuse magnetic scattering as a function of temperature in a positively exchange biased Co/FeF$_2$ bilayer and analyzed the data in the distorted wave Born approximation to obtain in-plane charge and magnetic correlation lengths associated with the Co and FeF$_2$ layers and estimate interfacial roughness. Tuning to the Fe and Co L$_3$ edges reveals significantly different temperature trends in these quantities in the antiferromagnetic and ferromagnetic layers, respectively. While the magnetic correlation length of the uncompensated interfacial spins in FeF$_2$ layer increase as temperature decreases, these quantities remain unchanged in the Co layer. Our results indicate that uncompensated Fe spins order within a range of few hundred nanometers in otherwise randomly distributed uncompensated magnetic moments, giving rise to spin clusters in the antiferromagnet whose size increase as the temperature decrease.

Quantitative statistical characterization of in-plane structural and magnetic feature is crucial for understanding the role of lateral magnetic heterogeneity in exchange coupled systems. The interfacial nature of the exchange bias problem immediately points to the important role played by various in-plane features with their associated length scales in the ferromagnet (F) and the antiferromagnet (AF) [1-3]. By controlling and tailoring the interfacial parameters over relevant length scales it is possible to strike a balance between exchange, Zeeman and disorder energy terms to obtain negative [4,5], positive [6], and even co-existing positive and negative exchange biased sub-loops [2,7]. Interfacial domain imaging clearly reveals that the F/AF interface is magnetically heterogeneous and that imprinting of the F domains on the antiferromagnet occurs [3, 9-11]. Depth resolved magnetization density profile studies performed using magnetic reflectivity have shown that the in-plane-averaged magnetization density across the F/AF interface is very different from the bulk of the thin film [12-14], with uncompensated AF spins extending throughout the AF layer and those within roughly 2 nm from the F layer free to rotate with the F and those farther from the interface fixed. However, many details of the in-plane spin behavior at the interface remain poorly understood. For example, how do the F and AF interface evolve laterally as a function of temperature? Do AF in-plane spins order? Can spin-blocks be detected? What happens to the magnetic interface above the Néel temperature?

A quantitative statistical description of the in-plane structure of a buried interface may be obtained using diffuse scattering in off-specular reflection geometry [15,16], which by employing resonant x-rays could be extended to study the F and AF interfacial spin structures [17-21]. In spite of its power and usefulness, rigorous analysis of magnetic diffuse scattering data is a non-trivial problem, and so examples in the literature are rare [22]. We address the question of in-plane structural and magnetic correlation lengths by performing a systematic temperature dependent resonant x-ray magnetic diffuse scattering study on an exchange biased Co/FeF$_2$ bilayer. We analyze the data in the Distorted Wave Born Approximation (DWBA), and determine quantitative values for element specific charge and magnetic correlation lengths. The results provide direct evidence of a frustrated magnetic interface and show that as a function of temperature the F and AF in-plane structure behaves very differently. This study shows that uncompensated spins of the AF at the F/AF interface give rise to a weak net moment, are correlated within a range of several hundred nanometers and the correlation length scale increases with decreasing temperature.

Diffuse scattering experiments were performed at beamline 4.0.2 at the Advanced Light Source of the Lawrence Berkeley National Laboratory using circularly polarized incident x-rays. The exchange biased Co/$FeF_2$ sample grown on an $MgF_2$ substrate investigated in this paper has been used previously in polarized neutron reflectivity measurements [14] and is nominally similar to the sample studied in a combined reflectivity study using resonant soft x-rays and neutrons [12]. The electron density profile of the sample was determined from non-resonant (Cu $K_\alpha$) x-ray reflectometry. The thickness (roughness) of the Co (Co/$FeF_2$) and $FeF_2$ ($FeF_2$/$MgF_2$) layers (interfaces) are 5.3±0.2 nm (0.9 ±0.1 nm) and 36.2 ±7.9 nm (0.4 ±0.1 nm) respectively. The sample was cooled to 15 K in a field of $H_{FC}$ = 1 T applied along the [001] $FeF_2$ easy axis to establish positive exchange bias [6]. The sample was field cycled three times at 15 K to eliminate any training effects. Diffuse scattering was measured at temperatures spanning the Néel temperature of bulk $FeF_2$ ( $T_N$ = 78 K) of 15 K, 55 K, 100 K, and 300 K in an applied field H = 1 T. The incident x-ray energy was tuned to either the Co or Fe $L_3$ edge with the energy slightly below the resonant peak to ensure that we know unambiguously the sign of scattering factors, as they vary across the resonant edge [23]. Rocking curves at the Co edge were taken at E = 775 eV and 2θ = 8 degrees ($q_z$ = 0.0547 Å$^{-1}$) while the corresponding values at the Fe edge were E = 706.6 eV ( $q_z$ = 0.0498 Å$^{-1}$).

Resonant diffuse scattering intensities were measured using circular polarization having opposite helicity, $I_+$ and $I_-$. It has been demonstrated theoretically and experimentally that the intensity difference ($I_+ - I_-$) results predominantly from charge-magnetic cross-correlations while the intensity sum ($I_+ + I_-$) results predominantly from charge-charge plus magnetic-magnetic correlations [17,18,20]. Analysis of the diffuse scattering data thus provides element specific structural insight into correlation lengths, interfacial roughness parameters, and roughness exponents that correspond to pure charge, pure magnetic, or mixed magnetic-charge structure. These correlation lengths represent the distance over which the in-plane chemical (charge) and magnetic structure remain unchanged. Structural correlation lengths are often associated with the characteristic length scale of interfacial roughness for films whose chemical composition is otherwise homogeneous, as we believe to be the case for this sample. In $I_+$ and $I_-$ we assume that only the 1$^{st}$ order magnetic term [18] contributes so that the predominant magnetic contributions come from variation in the net Co moment in the F layer and net uncompensated Fe moments in the AF layer. In-plane magnetic correlation lengths can thus be influenced both by the spatial distribution of magnetic species having uniform magnetization as well as magnetization variation of magnetic species having uniform spatial distribution.

Below we present the temperature dependence of $(I_+ + I_-)$ and $(I_+ - I_-)$ measured at the Co and Fe edges. We first discuss some general aspects of observed features in terms of magnetic and charge contributions. Subsequently we describe modeling of these data aimed at extracting in-plane charge and magnetic correlation lengths and their evolution with temperature in the Co and FeF$_2$ layers.

Diffuse scattering data are shown in Figure 1, where $I_+ + I_-$ and $I_+ - I_-$ are in the top and bottom rows, respectively, and resonant Co and Fe data are in the left and right panels, respectively. The different datasets are offset for clarity and plotted with the specular peaks truncated to highlight only the diffuse scattering in the tails of the specular beam. When tuned to the Co edge the resonant magnetic scattering from Fe is negligible, and vice versa. The signal at the Fe edge is weaker compared to Co because of relatively fewer number of uncompensated spin that give rise to a weak net moment. Nevertheless, the presence of magnetic diffuse scattering clearly shows the laterally heterogeneous distribution of the AF uncompensated moments. Diffuse magnetic scattering can arise from both the domains and spin disorder, which will manifest as magnetic roughness. Since at sufficiently high fields close to saturation, the typical size of magnetic domains is larger than the coherence length of the x-ray beam, the diffuse scattering from domains can be neglected. Through out the measured temperature range we have not observed features in the diffuse curves that can be ascribed to long range ordering of domains, rather the diffuse curves show a smooth decay as a function of in-plane **q** away from specular as would be obtained due to a randomly disordered surface [15].

The curves also show small humps that are more pronounced in the $(I_+ - I_-)$ channel. These are the Yoneda peaks; they do not represent structural or magnetic ordering peaks in the sample. The Yoneda peak occur because when the incident or exit angle equals the critical angle the total surface electric field reaches twice its value compared to the incident field thereby giving rise to enhanced diffuse scattering [15] as calculated in the distorted wave

Born approximation (DWBA). An interesting observation is that the Yoneda peak intensity associated with resonant scattering is more pronounced in the ($I_+$ - $I_-$) channel and increases as the temperature decreases. This is because the Yoneda scattering has a magnetic component and as the magnetization increases, the strength of the magnetic scattering increases, and hence a stronger magneto-optical effect whenever the incident or exit angle equals the critical angle.

We have developed analysis tools following the theoretical formalism of x-rays in the DWBA for resonant magnetic diffuse scattering as detailed in Ref [24]. We considered a model which is consistent with our earlier model used to fit resonant specular reflectivity data [12]. The model consists of a single layer of Co with the magnetic surface being almost the same as structural surface to reflect the saturation state of Co. The $FeF_2$ layer was divided into two layers: a thin region near interface of thickness ~2 nm which is an outcome of fitting result, and the remaining (bulk) slab of $FeF_2$. The total thickness was constrained to 36.2 nm. Dividing the Co layer into two layers shows no big difference in the fitting parameters compared to a single Co layer. We note that using the same model we were able to fit the specular reflectivity as well as the diffuse scattering data thereby showing the validity and accuracy of the model.

Each layer is described by thickness (taken as the same for charge and magnetic), roughness, scattering factor (real and imaginary part) for charge and magnetic, as well as the correlation length (in-plane and out-plane) and roughness exponent, together with the correlation length between charge and magnetic. The correlation function is given by,

$$C_{ll',nn'}(R) = \frac{\sigma_{l,n}\sigma_{l',n'}}{2}(e^{-(\frac{|R|}{\xi_{ll',n}})^{2h_{ll',n}}} + e^{-(\frac{|R|}{\xi_{ll',n'}})^{2h_{ll',n'}}})e^{\frac{-|z_n - z_{n'}|}{\xi_{\perp,ll'}}} \quad (1)$$

and represents the cross-correlation between $ll'$ for charge (cc), magnetic (mm), and charge-magnetic (cm) at interfaces $n$ and $n'$, $\sigma$ is the roughness, $h$ is the roughness exponent, $\xi$ and $\xi\perp$ are the in-plane and vertical correlation length, respectively. We find that for both F and AF the vertical correlation is larger than the film thickness. This suggests that the features along the depth of the sample are correlated and we have not included the vertical correlation length into the discussion.

To fit the diffuse scattering curves, we use structural thickness and roughness that were obtained by fitting non-resonant Cu K$_\alpha$ reflectometry data, and the refractive indices of the non-resonant layers, which were obtained from standard tables [25]. The charge-charge correlation length and roughness exponent were obtained by fitting diffuse data at 650 eV, which is far away from the resonant edges for either Co or Fe where the diffuse intensity is predominantly due to charge scattering with negligible magnetic contribution. The charge related parameters were constrained to remain fixed. The (I$_+$ + I$_-$) and (I$_+$ - I$_-$) data were simultaneously fitted by varying the magnetic roughness ($\sigma_m$), magnetic-magnetic ($\xi_{mm}$), and charge-magnetic ($\xi_{cm}$) correlation lengths and the real and imaginary parts of the complex refractive index for the resonant layer. As shown in Fig. 1, a reasonably good fit has been obtained for most features in the diffuse scattering curve both in the sum (I$_+$ + I$_-$) and difference (I$_+$ - I$_-$) channel including the Yoneda peaks.

The behavior of the magnetic roughness ($\sigma_m$) as a function of temperature is shown in Fig. 2. The magnetic roughness of the Co layer is fixed to equal the charge roughness at 0.9 nm for all temperatures. Thus the magnetic surface is made to follow the structural surface, and magnetic diffuse scattering is due to the spin disorder at the F/AF interface whose origin lies at the structural roughness. In contrast the magnetic roughness of the Fe layer does show temperature behavior. First, the magnetic roughness of the top Fe layer (1.0 nm) is larger than the layer below (0.4 nm). Recalling that the top FeF$_2$ layer is 2 nm in thickness, a 1 nm roughness means that a significant portion of the top Fe layer is significantly frustrated. Thus, for the Fe layer, the spins are more disordered near the F/AF interface than the spins away from interface. Second, the magnetic roughness of the top Fe layer decreases from 1.4 nm at 300 K to 1.0 nm at 15 K which means the magnetic surface is smoother at low temperatures enabling stronger interaction between F and AF across the interface [26].

Figure 3 shows the temperature dependence of the correlation lengths at the Co (a) and Fe (b) edge. The charge-charge correlation length $\xi_{cc}$ for both Co and Fe at the F/AF interface is fixed because the structural roughness depends on the sample growth parameters and is temperature independent. At 300 K the magnetic correlation length $\xi_{mm}$ for Co (F) is 60 (±4) nm and Fe (AF) is 13 (±5) nm. As has been reported [27, 28], the exchange induced uniaxial anisotropy for Co/FeF$_2$ persists well above T$_N$ up to room temperature, that gives rise to nanoscopic regions where the F is coupled to AF. In these local regions the F polarizes the AF (which is now disordered) spins and some of the F in-plane structure may get imprinted on the AF [8, 29].

As the temperature is decreased, passing through $T_N$ of $FeF_2$ (78 K), $\xi_{mm}$ due to the uncompensated moments in the AF continue to increase, while the ferromagnetic correlation lengths remain essentially unchanged. At 300 K, the ratio of $\xi_{mm}$ in the Co and FeF2 is 3:1. At 15 K the ratio has reversed to approximately 1:3, indicating that the imprinting by the F at 300 K is quickly dominated by effects internal to the AF. $\xi_{mm}$ for Co show almost no change as a function of temperature, which is quite reasonable because almost nothing changes with temperature for Co below Curie temperature and at saturation field. It is interesting to note that $\xi_{cc}$ is slightly larger than $\xi_{mm}$ (80 (±6) nm to 60 (±4) nm) for the F. However, $\xi_{mm}$ and $\xi_{cm}$ have similar values (with $\xi_{cm}$ intermediate between $\xi_{cc}$ and $\xi_{mm}$) that shows coupling of the magnetic with charge features. We therefore conclude that a 1 T applied magnetic field that should nominally saturate the F is not enough to saturate the interfacial F spins that are closely coupled to the surface structural disorder. In addition, since the measurement temperature is far below $T_C$, the interfacial Co spins are frozen [30]. This result is consistent with our previous specular reflectivity studies where we observed that at the biased condition the magnetic F/AF interface is broader than the structural one [12-14].

In contrast, at the Fe edge $\xi_{mm}$ increase faster than $\xi_{cm}$. At 300 K both $\xi_{mm}$ and $\xi_{cm}$ are significantly lower than $\xi_{cc}$ indicative of a higher degree of interfacial spin frustration than surface charge or structural disorder. Below 100 K $\xi_{mm}$ and $\xi_{cm}$ is larger than $\xi_{cc}$, and at 15 K $\xi_{mm}$ is more than twice the value of $\xi_{cc}$. The low temperature in-plane magnetic structure is therefore predominantly unrelated to structural disorder although magnetic features coupled with charge feature still exists as evidenced by the growth of $\xi_{cm}$. For T ≤ 100 K we find that both $\xi_{mm}$ and $\xi_{cm}$ of the AF is larger than the corresponding lengths in the F. Thus, below $T_N$ the Fe uncompensated interfacial spins develops their own in-plane structure, although some imprinted structure from the F may persist till low temperatures.

Apart from the structural roughness that is a source of spin disorder, the exchange-induced uniaxial anisotropy and Zeeman effect, both having temperature dependences, are the most likely origin of frustrated uncompensated AF spins at the interface. Our diffuse scattering results indicate that the uncompensated spins in the AF are ordered within a range of about 100 - 200 nm, which we define as a short-range order, in the plane of the F/AF interface in an otherwise randomly distributed uncompensated spin structure. The short-range ordered spins give rise to islands or spin clusters, the size of which is given by the corresponding correlation length. As the temperature decreases, more spins get aligned along the easy axis of FeF$_2$ resulting in an increase of the size of the spin cluster, as seen by an increase in correlation length. The overall effect of the increase in the correlation length of the uncompensated AF moments and decrease of magnetic roughness is to create a smoother and more stable magnetic interface that favors a stronger coupling between the F and the AF [26,31]. While it has long been conjectured and numerous experiments indirectly indicate that the AF interface is responsible, this is the first evidence of temperature and length scale evolution of uncompensated in-plane spin structure of the AF at the F/AF interface showing that it is indeed correlated with the exchange bias [6].

Finally we get some insight into the evolution of the line shape of magnetic diffuse scattering as a function of magnetic-magnetic correlation length. For this we set the charge-related parameters (charge scattering factors and charge roughness) to zero, and simulate diffuse scattering curves that are purely magnetic in origin. As we noted before, in Born Approximation the $I_+ - I_-$ signal is related to charge-magnetic interference term. But in the DWBA the difference signal has additional terms due to pure magnetic scattering. The magnetic diffuse scattering at the Co and Fe edges as a function of the in-plane magnetic correlation length are shown in Fig. 4. Several interesting observations could be made which might help in interpreting diffuse scattering signal. First, we note that there are two peaks, assigned as magnetic Yoneda peaks, in each rocking curves. The magnetic susceptibility (or the refraction index) is different for the right and left circularly polarized x-ray resulting in slightly different critical angles. Hence $I_+$ and $I_-$ have different magnetic Yoneda peaks. Second, the lineshape for Co and Fe are quite different. This is because of the layer positions, thickness and layer magnetization. Another interesting observation in Fig. 4 is that the intensity dependence of the magnetic in-plane correlation length displays a maximum around 100 nm and 250 nm at the Co and Fe edges, respectively. At small correlation lengths, few spins are correlated and this leads to the weak diffuse scattering. At very large correlation lengths, the correlation function becomes almost flat with very smooth roughness, which also results in weak diffuse intensity. We also find that the diffuse signal in the ($I_+ - I_-$) channel can sometime reverse sign (see Fig 4(a)). This means that the sign of the hysteresis loops measured in diffuse condition depend not only on the magnetization direction but also on the specific Fourier component. Thus only by measuring a series of hysteresis loops as a function of wave-vector transfer can provide information about q-resolved in-plane magnetization.

In conclusion, soft x-ray resonant magnetic diffuse scattering can provide quantitative statistical description of buried interface in the nanometer length scale. We show that for a positively exchange biased $Co/FeF_2$ sample, imprinting of the F in-plane structure occurs above the Neel temperature, but the AF nevertheless develops its own in-plane structure when temperature is decreased below $T_N$. The magnetic-related correlation lengths remain unchanged for the F, while these values in the AF increase as the temperature decreases. The increase of correlation length in the AF implies less disorder that results in an effectively stronger exchange coupling between F and AF and a higher exchange bias magnitude. Thus disorder engineering can be used to tailor exchange bias.

This work at ALS/LBNL was supported by the Director, Office of Science, Office of Basic Energy Sciences, of the U.S. Department of Energy under Contract No. DE-AC02-05CH11231. S.K. Sinha and X. Lu were supported by DOE-BES Grant Number:DE-SC0003678. I.K.S and M.E were supported by DOE-BES Grant Number DE-FG03-87ER-45332.

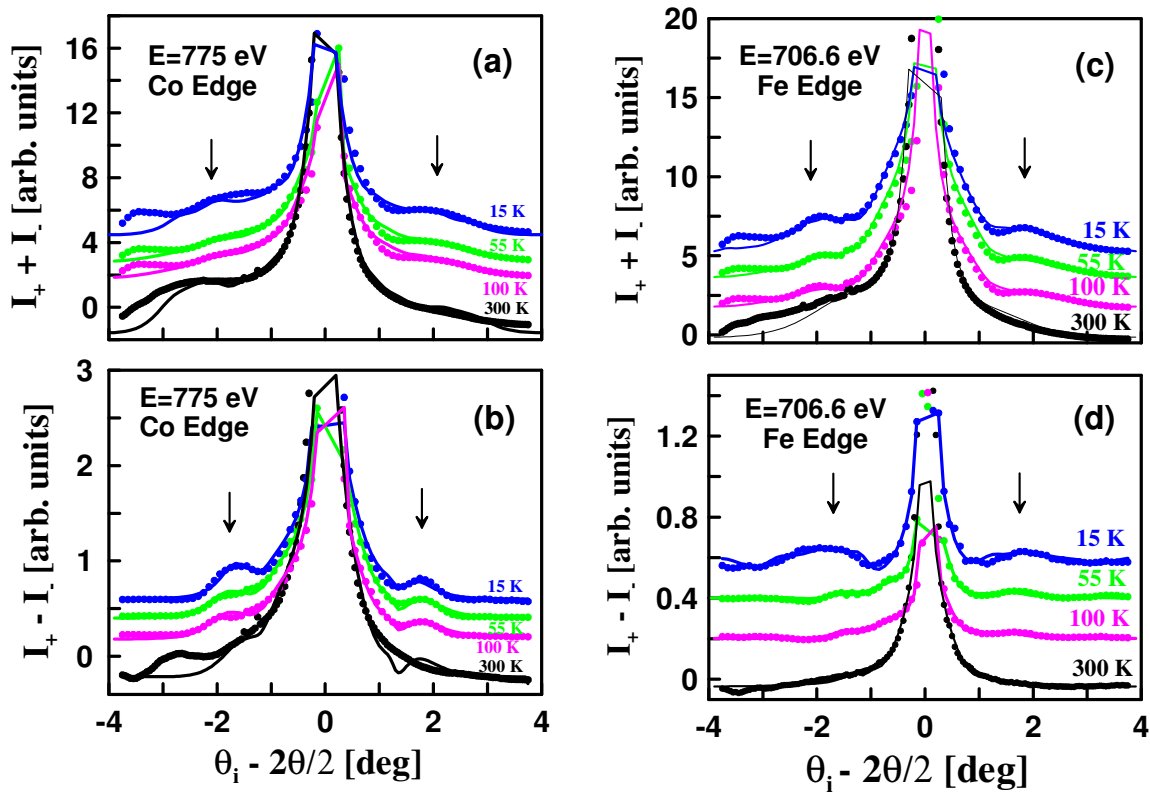

**Figure 1:** Diffuse scattering intensities at Co $L_3$ edge showing $(I_+ + I_-)$ (a) and $(I_+ - I_-)$ (b) and Fe $L_3$ edge $(I_+ + I_-)$ (c) and $(I_+ - I_-)$ (d) at temperatures 15 K, 55 K, 100 K and 300 K, where $I_+$ and $I_-$ are the scattered intensity with right- and left-circularly polarized incident beam. All the measurements were made at an applied magnetic field of 1 T. The dots are the data and solid lines are the fits. For clarity different data sets are vertically shifted with the specular peak truncated to highlight the diffuse tails. Arrows mark Yoneda peaks as determined from modeling as discussed in text.

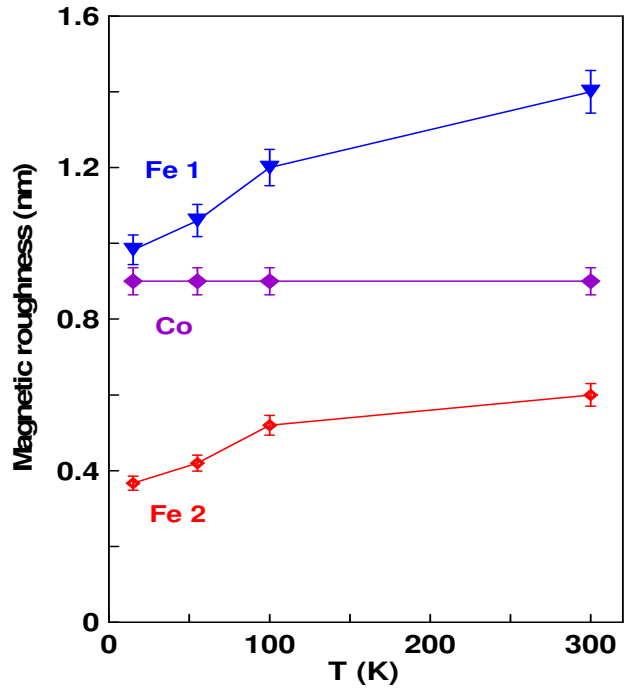

**Figure. 2:** Magnetic roughness at the Co and Fe $L_3$ edge. The magnetic roughness of the F (Co) layer is fixed, It decreases for the AF (Fe) as a function of temperature. Fe 1 is the topmost FeF$_2$ layer near the F/AF interface and Fe 2 is the remainder of the FeF$_2$ layer. Solid lines are guide to the eyes.

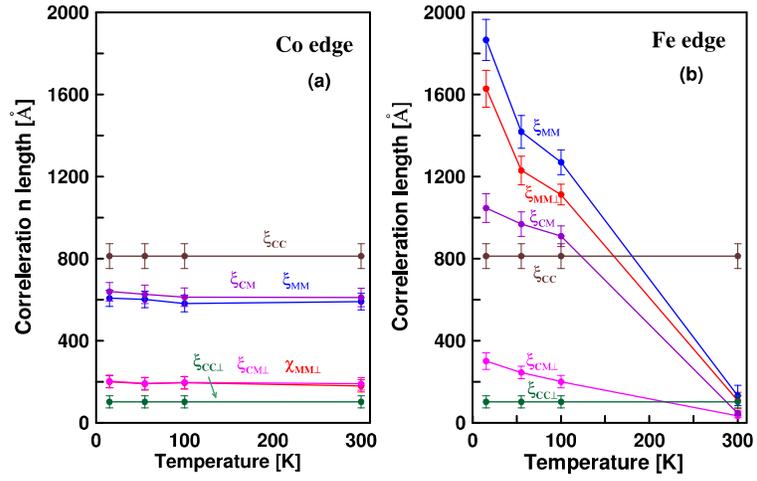

**Figure 3:** The correlation lengths $\xi$ as a function of temperature at Co edge (a) and at Fe edge (b). CC, CM and MM represent for charge-charge, charge-magnetic and magnetic-magnetic correlation, respectively. The dots are the fitting results; solid lines are guide to the eyes.

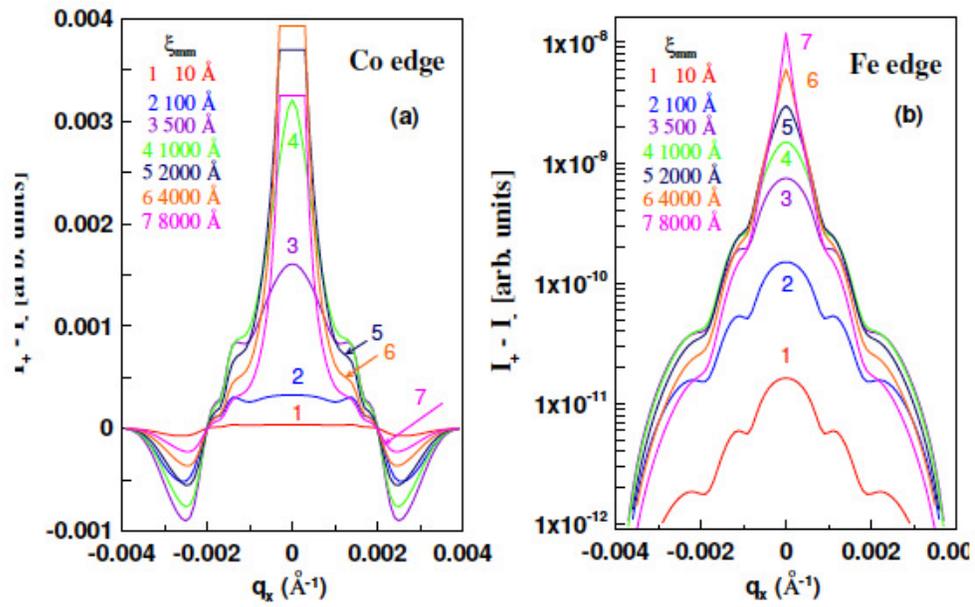

**Figure 4**. Calculated diffuse scattering intensity ($I_+ - I_-$) from *pure* magnetic scattering at Co edge (a) and Fe edge (b), as a function of the in-plane magnetic correlation length.